\documentclass[article,twocolumn]{revtex4-1}
\usepackage[T1]{fontenc} 

\usepackage{amsfonts}
\usepackage{amsmath}
\usepackage{amssymb}
\usepackage{graphicx}
\usepackage{subfigure}
\usepackage[utf8x]{inputenc}
\usepackage{color}
\usepackage{colortbl}
\usepackage{bm}
\usepackage{rotating}
\usepackage{soul}

\begin{document}

\title{Lasing Conditions of Transverse Electromagnetic Modes in Metallic-Coated Micro- and Nanotubes}

\author{Nicol\'as Passarelli}
\affiliation{Instituto de F\'isica Enrique Gaviola (IFEG-CONICET) and Facultad de Ciencias Qu\'imicas, Universidad Nacional de C\'ordoba, Ciudad Universitaria, C\'ordoba 5000, Argentina}
\author{Ra\'ul Bustos-Mar\'un}
\affiliation{Instituto de F\'isica Enrique Gaviola (IFEG-CONICET) and Facultad de Ciencias Qu\'imicas, Universidad Nacional de C\'ordoba, Ciudad Universitaria, C\'ordoba 5000, Argentina}
\email{rbustos@famaf.unc.edu.ar}
\author{Ricardo Depine}
\affiliation{{Grupo de Electromagnetismo Aplicado, Departamento de F\'{\i}sica, FCEN, Universidad de Buenos Aires and IFIBA, Consejo Nacional de Investigaciones Cient\'{\i}ficas y T\'{e}cnicas (CONICET), Ciudad Universitaria, Pabell\'{o}n I, Buenos Aires C1428EHA, Argentina}}

\begin{abstract}
In this work, we study the lasing conditions of the transverse electric (TE) modes  of micro- and nanotubes coated internally with a thin metallic layer.
This geometry may tackle some of the problems of nanolasers and spasers as it allows the recycling of the active medium while providing a tunable plasmonic cavity.
The system presents two types of TE modes: cavity modes (CMs) and whispering-gallery modes (WGMs).
On the one hand, we show that the lasing of WGM is only possible in nanoscale tubes.
On the other hand, for tubes of some micrometers of diameter, we found that the system presents a large number of CMs with lasing frequencies within the visible and near-infrared spectrum and very low gain thresholds. Moreover, the lasing frequencies of CMs can be accurately described by a simple one-parameter model.
Our results may be useful in the design of micro- and nanolasers for ``lab-on-chip'' devices, ultra-dense data storage, nanolithography, or sensing.
\end{abstract}

\maketitle

\section{Introduction}

Micro- and nanolasers (lasers of micro- or nanoscale dimensions)\cite{mccall1992,painter1999,johnson2001} and spasers
(a type of laser that confines light at subwavelength scales by exciting surface plasmon polaritons)
\cite{deeb2017plasmon,bergman2003} can be used to provide a controllable source of ``on-demand'' electromagnetic fields at very small size scales\cite{wang2017structural}.
These devices can find numerous applications for enhanced spectroscopies, as a way of counteracting the effect of optical losses in different plasmonic devices, or directly as a source of radiation for ``lab-on-chip'' devices, ultra-dense data storage, or nanolithography \cite{berini2012surface,hill2014advances,torma2015,amendola2017,ma2018}.
Different forms of nanolasers and spasers have been extensively discussed in the literature over the last decades, both from the theoretical\cite{
bergman2003,liu2017open,Warnakula,Huang2014Nonlocal,
Parfenyev2014Surface,Hapuarachchi2018Exciton,
Zhong2013all-analytical,ma2018}
and  experimental
\cite{mccall1992,painter1999,johnson2001,
ma2018,Ambati2008Observation,noginov2009demonstration,ma2014explosives,kiraz2015optofluidic,lai2013near,chandrahalim2015monolithic,zapf2017dynamical,lin2019broadband} point of view.
There has been studied a wide variety of structures and materials/metamaterials for supporting the spaser/nanolaser as well as different compounds to act as the optical-gain medium. This includes cavities in thin films, different forms of waveguides made of lithographic or nano-printed surfaces, graphene sheets, carbon nanotubes, nanoparticle or quantum dot arrangements, etc. \cite{deeb2017plasmon,wang2017structural}.

There are, in general, two main problems involved in the design of spasers and nanolasers.
One is the need for a good matching between the plasmonic resonances (for spasers) or the cavity resonaces (for nanolasers) and the emission cross-section of the active medium.
The other main problem is photobleaching, or the irreversible photo-degradation of the molecules\cite{chua2014modeling} or quantum dots\cite{kiraz2015optofluidic} that form the active media.
Other features are also desirable, such as large volumes of hot-spots\cite{zhao2014monolithic}, to allow the study of large molecules or the presence of several resonances that can be used for lasing with different active media\cite{lin2019broadband}.
The ideal design of a nanolaser or spaser should not only solve the photobleaching problem but also offer some flexibility to finely tune the frequencies of resonances\cite{lin2019broadband,zapf2017dynamical,li2018excitonic} while providing large volumes of hot spots and low gain thresholds.

\begin{figure}[t]
\includegraphics[width=3.0 in]{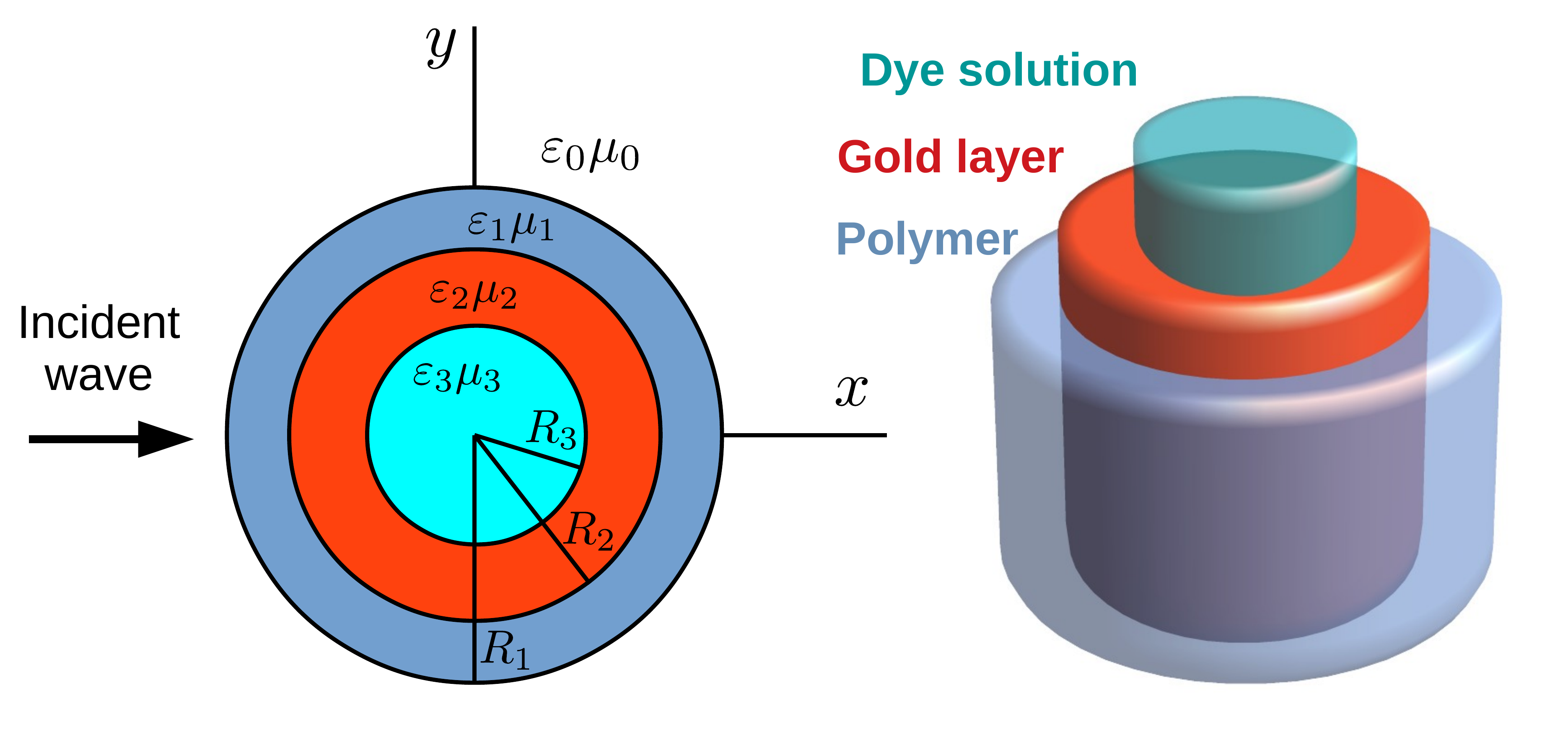}
\caption{Scheme of the system treated. An infinite multiwall cylinder where the core allows the circulation of a solution with the dye (active medium), the intermediate wall provides the plasmonic cavity, and the outer wall gives the physical support to the system. The ``incident wave'' is just used to numerically find the poles of the scattering cross-section. See text for more details.}
\label{fig:1}
\end{figure}
In this work, we study the lasing conditions of a geometry that seems promising to meet all the above requirements.
The proposed geometry, see Fig. \ref{fig:1}, allows the recycling of the active medium (owing to microfluidics\cite{chandrahalim2015monolithic}), provides a tunable resonant cavity (by modifying the dielectric constant of the solution of the core) physically separates the active medium from the external medium (where analytes may be placed for instance), and presents large volumes of hot spots, low gain threshold, and a large number of resonances within the visible and near-infrared spectrum.

The work is organized as follows.
In the General remarks section, we discuss some generalities of the studied system.
In the Scattering Problem section, we explain how to solve the scattering problem in the linear-response regime.
There, we give the equations of the electromagnetic fields in terms of cylindrical harmonics, the boundary conditions, the expressions for the cross sections, and the relation between the scattering cross sections and near fields at the lasing condition.
In section ``Modeling the optical active medium'', we discuss the different approximations used to describe the active medium and how we obtained the parameters of the dyes used. 
In the Lasing conditions section, we discuss the lasing conditions in terms of lasing frequencies and gain thresholds, providing also some approximations for their estimation. 
In the Results and Discussion section, we show the main results of the work and summarize them in the Conclusions section.

\section{Theory}
\textbf{General Remarks.\label{sec:remarks}}
The theoretical treatment of the system is sustained in two major axes.
The first one is an extension of well-known analytical solution of an infinite two-wall
concentric cylinder\cite{shah1970,kerker1961, hulst1981,bohren2008absorption} to three-wall concentric cylinders, including the expressions for the scattering and extinction cross sections.
The second axis is the modeling of the active medium. We use two different models: a wideband approach, which neglects the frequency dependence of the active medium,  and a Lorentz-like model for the frequency dependence of the active medium.

Although we performed several calculations with different geometries, most of the work is focused on one representative example.
This geometry consists of three concentric cylinders, (see Figure \ref{fig:1}).
The internal core of 2000 nm of radius consists of a dielectric solvent and a dissolved dye.
The inner wall of the tube is a layer of 25  nm of a plasmonic material (gold), and the outer one, of 475  nm, is a dielectric (polymer) that provides mechanical support to the cavity and separates it from the external medium.

Because of its sizes, the system lays between the micro-laser and nanolaser categories. Furthermore, because of the metallic coating, the lasing of some resonances turn the system into a spaser. For simplicity, throughout this paper, we will use the generic term nanolaser to refer to the system but keeping in mind these considerations.

The geometry was inspired by recent works that present polymer microtubes suitable for fluidics \cite{yarin2007,dror2007,vandersarl2011nanostraws,kim2013cavity,kiraz2015optofluidic}
The fact that the system allows the circulation of a fluid through it can be used for recycling the active media and for depositing the metal layer by chemical methods.\cite{hu2008novel,formanek2006selective,demoustier2001preparation}
The recycling of the active media may help counteract the photobleaching of the dye molecules acting as the optically active medium.
The intermediate layer provides the plasmonic cavity that will control the resonance frequencies and the outer layer gives the physical support to the system.
In this way, if, for example, the nanolaser is used for sensing purposes, the analyte is physically separated from the active medium.
The latter not only avoids undesired chemical interferences but can also be used to control the resonances of the nanolaser by tunning the dielectric constant of the inner cylinder.

Similar geometries have been studied before, including wires, tubes, pores, and fibers or capillaries.\cite{garreau2014,hill2007,nezha2010,ding2013,kim2013cavity,kiraz2015optofluidic}
The sizes of these systems ranged from tens of nanometers up to hundreds of micrometers of radius.
The geometries within the nanometers scale have been studied theoretically for perpendicular illumination, which excites the TE modes of lasing, but generally within quasi-static approximations.
The larger geometries have been studied mainly in the context of low loss communication, where the longitudinal modes of lasing are the ``important'' ones.\cite{pan2005,shevchenko2005,moon2004,parola2016,saleh2012,wang2015,wang2018} The resonances in such a case can be thought as Fabry P\'erot modes\cite{bordo2010} for finite systems, or they can be treated as dispersion relations for infinite cylinders.\cite{tong2004}

Regarding the feasibility of the proposed geometry, modern techniques of fabrication show that tubes of a few micrometers, suitable for microfluidics, can be produced \cite{huang2017,persano2015,gu2010,camposeo2009,ishii2016}
and it is even possible to make them into different materials
\cite{tang2009,vasdekis2007,dror2007,yarin2007,goldberger2006}.
On the other hand, coating of surfaces with very thin layers of noble metals has been employed in several experimental works using different techniques.\cite{wirtz2002template,wirtz2003template}

The incident wave, shown in Figure \ref{fig:1}, is used in the calculations to find the poles of the scattering cross section of the system, which leads us to the lasing conditions. We will only consider the case of electric fields perpendicular to the $z$ axis of the cylinders as we are studying nonmagnetic materials. For this polarization, the resonances will be found when the perpendicular component of the incident wave matches the resonant condition, and this is why we need only to consider the perpendicular incidence.

The dielectric constant of gold as a function of the wavelength was taken from ref  \cite{johnson1972}. The dielectric constant of the polymer was considered constant with $\varepsilon_{1}=2.22$ \cite{sultanova2009}. The theoretical treatment of the dielectric constant of the active medium is discussed in details in the Modeling the Optical Active Medium section.

\textbf{Scattering Problem.\label{sec:linear_response}}
The analytical solution for the scattering of electromagnetic incident plane waves  perpendicularly on a homogeneous circular cylinder of infinite length is well known \cite{hulst1981}. In this case, and because of the translational invariance of the geometry, the solution of the rigorous electromagnetic vectorial problem can be reduced to the treatment of two independent scalar problems, corresponding to the following two basic polarization modes in which the polarization of the interior and scattered fields maintain the polarization of the incident wave: 
 (i) $s$ (or TM) polarization, when the incident electric field is directed along the cylinder axis, and 
 (ii) $p$ (or TE) polarization, when the incident magnetic field is directed along the cylinder axis
In the first case, it is clear that the electric field can only induce electric currents directed along the cylinder axis but not along the azimuthal direction, thus preventing the existence of localized surface plasmons in metallic cylinders for $s$ (or TM) polarization. For $p$ polarization, on the other hand, the electric field, contained in the main section of the cylinder, can induce azimuthal electric currents, which  under appropriate conditions, give place to localized surface plasmon resonances. 

For multilayered cylinders with an arbitrary number of layers \cite{kerker1961,shah1970,gurwic1999}, similar conclusions can be obtained: the solution of the rigorous electromagnetic vectorial problem can be separated into  two independent ($s$ and $p$) scalar polarization modes, and when metallic layers are involved, localized surface plasmon resonances occur for $p$, but not for $s$, incident polarizations. Taking into account that we are interested in the TE modes of nanolasers, in what follows, we restrict our attention just to the case of $p$ polarization.

To model the linear response of our system, we consider an infinitely long circular cylinder (the core) covered with two concentric layers of homogeneous materials and illuminated from a semi-infinite  ambient medium. We assume that all the materials involved are isotropic and linear and characterized by a dielectric constant $\varepsilon_j$ and a magnetic permeability $\mu_j$ ($j=0, 1, 2, 3$), where $j=0$ corresponds to the ambient medium and $j=3$ corresponds to the core (see Figure \ref{fig:1}).
The Gaussian system of units is used and an $\exp(+i\omega t)$ time dependence is implicit throughout the paper, with $\omega$ being the angular frequency, $c$ being the speed of light in vacuum, $t$  being the time, and $i=\sqrt{-1}$.

\textbf{Fields Representation.\label{sec:fields_representation}}
Using polar coordinates ($r, \theta$), where $\theta$ is the angle respect to the $x$ axis of a vector contained in the $x-y$ plane of modulus $r$, all the fields in the $p$ polarization case can be written in terms of $F_j (r,\theta)$,  the non-zero  component of the total magnetic field along the axis of the cylinder and evaluated in region $j$. 
According to Maxwell's equations, $F_j (r,\theta)$ must satisfy Helmholtz equations in each region 
\begin{equation}
(\nabla^2+k_j^2)F_j=0 \,,
\end{equation}
with $k_j=\omega m_j/c$ and $m_j=(\varepsilon_j\,\mu_j)^{1/2}$ being the refractive index of medium $j$. 
Following the usual separation of variables approach, $F_j (r,\theta)$ can be represented with the following multipole expansions 
\begin{eqnarray}
F_j(r,\theta)=\sum_{n=-\infty}^{\infty} [a_{jn} J_n + 
A_{jn}\, H_{n} ] \,{\rm e}^{i n \theta} \, \label{eqn:3Cz2} 
\end{eqnarray}
where $a_{jn}$ and $A_{jn}$ ($j=0,\ldots 3$) are complex amplitudes,  $J_{n} (k_j r)$ is the $n$-th Bessel function of the first kind, and $H_{n} (k_j r)$ is the $n$-th Hankel function of the second kind, with asymptotic behavior 
for large argument ($|x|\rightarrow \infty$) 
\begin{equation}
H_{n} (x) \longrightarrow  \sqrt{\frac{2}{\pi x}} \,{\rm e}^{\,-i (x -n \pi/2 - \pi/4)} \,.
\label{eqn:3Casintot}
\end{equation}
The amplitudes $a_{0n}$ in the medium of incidence are determined by the incident wave. 
Assuming, without loss of generality, that a unit-amplitude incident plane wave propagating along the x-axis, and using the Jacobi Anger expansion\cite{dattoli1996theory}
\begin{equation}
{\rm e}^{-i k_0 r \cos \theta} 
= \sum_{n=-\infty}^{\infty} (-i)^n J_n (k_0 r) \,{\rm e}^{i n \theta} 
\end{equation}
the amplitudes $a_{0n}$ are given by
\begin{equation}
a_{0n}=(-i)^n \,.
\end{equation}
For physical reasons, Hankel functions of the second kind, singular at the origin, are not allowed in the field representation of the core region. Thus 
\begin{equation}
A_{3n}=0. 
\end{equation}
Therefore, to evaluate the magnetic field $\vec H_j = F_j \,\hat z$ everywhere with \eqref{eqn:3Cz2}, a sextuple infinity of unknown amplitudes, namely, $A_{0n}$,$ a_{1n}$, $A_{1n}$,$ a_{2n}$,$ A_{2n}$and $ a_{3n}$, must be  found. Finally, the electric field $\vec E_j$ in $p$ polarization in terms of $F_j (r,\theta)$ can be obtained from Amp\`ere's law
\begin{equation}
\nabla \times (F_j \,\hat z)  = i\,\frac{\omega}{c} \varepsilon_j\, \vec E_j \,.  \label{eqn:rotorp}
\end{equation}

\textbf{Boundary Conditions.\label{sec:boundary_conditions}}
At the boundary $r=R_j$ ($j=1,\ldots 3$) between two homogeneous regions, the continuity of the tangential components of the electric and magnetic fields must be fulfilled, and these conditions are equivalent to the continuity of $F_j$ and  $\frac{1}{\varepsilon_j} \frac{\partial F_j}{\partial r}$. 
These boundary conditions provide a system of six linear equations for the six unknown amplitudes $A_{0n}, a_{1n}, A_{1n}, a_{2n}, A_{2n}$ and $a_{3n}$. Due to the rotational symmetry of the system, the boundary conditions do not couple amplitudes with different multipole order $n$. 
The continuity of the $z$ component of the magnetic field gives equations 

\begin{widetext}
\begin{eqnarray}
a_{3n} J_n (k_3R_3)  = 
a_{2n} J_n (k_2R_3) + A_{2n} \, H_{n} (k_2R_3)  \label{eqn:3CFconti1}\\
a_{2n} J_n (k_2R_2) + A_{2n} \, H_{n} (k_2R_2)  = 
a_{1n} J_n (k_1R_2) + A_{1n} \, H_{n} (k_1 R_2)  \label{eqn:3CFconti2}\\
a_{1n} J_n (k_1 R_1) + A_{1n} \, H_{n} (k_1 R_1)   = 
(- i)^n J_n (k_0 R_1) + A_{0n} \, H_{n} (k_0 R_1)  \label{eqn:3CFconti3}
\end{eqnarray}
whereas the continuity of the tangential component of the electric field gives the equations 
\begin{eqnarray}
\frac{k_3}{\varepsilon_3} a_{3n} J_n' (k_3R_3)  = 
\frac{k_2}{\varepsilon_2} [\,
a_{2n} J_n' (k_2R_3) + A_{2n} \, H_{n}' (k_2R_3) \,]
\label{eqn:3CFconti4}\\
\frac{k_2}{\varepsilon_2} [\, 
a_{2n} J_n' (k_2R_2)  + A_{2n} \, H_{n}' (k_2R_2)   \,]= 
\frac{k_1}{\varepsilon_1} [\,
a_{1n} J_n' (k_1R_2)  + A_{1n} \, H_{n}' (k_1 R_2)   \,]
\label{eqn:3CFconti5}\\
\frac{k_1}{\varepsilon_1} [\, 
a_{1n} J_n' (k_1 R_1)  + A_{1n} \, H_{n}' (k_1 R_1)    \,] = 
\frac{k_0}{\varepsilon_0} [\,
(-i)^n J_n' (k_0 R_1)  + A_{0n} \, H_{n}' (k_0 R_1)  \, ]
\label{eqn:3CFconti6}
\end{eqnarray}
where primes denote derivatives of the Bessel and Hankel functions with respect to their arguments. 
The above equations can be written in the matrix form
\begin{equation}
\mathbb{M} \vec{X} = \vec{B} \label{eq:Matrix}
\end{equation}
 with $\vec{X} = \left [ A_{0n},a_{1n},A_{1n},a_{2n},A_{2n},a_{3n} \right ]^T$, 
$\vec{B} = (-i)^n \left [ J_n (k_0 R_1),\frac{k_0}{\varepsilon_0} J_n' (k_0 R_1),0,0,0,0 \right ]^T$, and
\begin{equation}
\hspace{-1.3cm}
\mathbb{M} = 
\begin{bmatrix} 
-H_{n} (k_0 R_1)                                              &  J_n (k_1 R_1)                                 &   H_{n} (k_1R_1)     &   0            &  0  &  0  \\
-\frac{k_0}{\varepsilon_0} H_{n}' (k_0 R_1) &  \frac{k_1}{\varepsilon_1} J_n' (k_1R_1)  &   \frac{k_1}{\varepsilon_1}  H_{n} (k_1R_1)    &  0  &  0 &  0  \\
0   & J_n (k_1R_2)  &  H_{n} (k_1 R_2)        & -J_n (k_2R_2)  & -H_{n} (k_2R_2) &  0 \\
0   &\frac{k_1}{\varepsilon_1} J_n' (k_1R_2)  &  \frac{k_1}{\varepsilon_1}H_{n}' (k_1 R_2)        & -\frac{k_2}{\varepsilon_2}J_n' (k_2R_2)  & -\frac{k_2}{\varepsilon_2}H_{n}' (k_2R_2) &  0 \\
0   &   0  &  0      &   J_n (k_2R_3)   &  H_{n} (k_2R_3)  & -J_n (k_3R_3)\\                                              
0   &   0  &  0      &  \frac{k_2}{\varepsilon_2} J_n' (k_2R_3)   &  \frac{k_2}{\varepsilon_2} H_{n}' (k_2R_3)  & -\frac{k_3}{\varepsilon_3} J_n (k_3R_3)
\end{bmatrix} \notag
\end{equation}
\end{widetext}

\textbf{Far Fields, Near Fields, and the Lasing Thresholds.\label{sec:far_field}}
The coefficients $A_{0n}$ determines the amplitude of the scattered wave and can be used to calculate the cross sections per unit length of the cylinder

\begin{equation}
C_{sca}=\frac{4\varepsilon_0^{2}}{k_0}\sum_{n=-\infty}^{\infty}|A_{0n}|^{2} \label{eq:C_sca}
\end{equation}
\begin{equation}
C_{ext}=\frac{4}{\varepsilon_0^{2}k_0}\sum_{n=-\infty}^{\infty}\mathrm{Re}\{A_{0n}\}
\end{equation}
where $C_{sca}$ and $C_{ext}$ are the scattering and extinction cross sections, respectively.

The poles of $C_{sca}$ and $C_{ext}$ for each order $n$, can be found as the points in the parameter's space where the determinant of the matrix $\mathbb{M}$ in eq \ref{eq:Matrix} cancels. 
There, the fields associated with a given order $n$ diverge. This condition corresponds to resonances with compensated optical losses, and they are identified with the lasing threshold.

As we have already reported for spheres,\cite{passarelli2016} there is a proportionality relation between the near and far field at the lasing condition.
This can readily be understood by considering the following reasons.
When the determinant of $\mathbb{M}$ approaches zero, all the coefficients of modes $n$ and $-n$ goes to infinity as $\vec{X}=\mathbb{M}^{-1} \vec{\mathbb{B}}$.
Let us recall that as $j_{-n}=(-1)^n j_n$ and $H_{-n}=(-1)^n H_n$, the coefficients $a_{j,n}$ and $a_{j,-n}$ should be equal.
Then, at the lasing condition, the summation in eq \ref{eqn:3Cz2} can be replaced by the summation of only two terms, $F_n$ and $F_{-n}$ (or $F_0$ for $n=0$). The same can be done to the cross sections $C_{sca}$ and $C_{ext}$, and the electric fields $\vec{E}_j$.

According to eq \ref{eqn:rotorp}, the component of the electric field $E$ can be written in term of the $F$ function as
\begin{equation}
 E_{r}=\frac{-inF}{r}\quad E_{\theta}=-\frac{\partial F}{\partial r}.
\end{equation}
Then, $\left|\vec{E}_{0n}\right|^{2}$, the square modulus of the scattered electric field, at the lasing condition of modes $n$ and $-n$, results in
\begin{equation}
\left|\vec{E}_{0n}\right|^{2} = \left |A_{0n} \right |^{2} \left |\vec{M}_n \right |^2	\label{eq:campocev-1}
\end{equation}
where 
\begin{eqnarray}
\left |\vec{M}_n \right |^2 & = & k_{0}^{2}
\left(\left |H_{n-1} \right |^{2}+ \left |H_{n+1} \right |^{2}\right) \notag \\
&& \times 
\begin{cases}
2\cos^{2}\left(n\theta\right) & n\:\mathrm{even}\\
2\sin^{2}\left(n\theta\right) & n\:\mathrm{odd}\\
\frac{1}{2} & n=0
\end{cases}  \notag
\end{eqnarray}
We have used above the recurrent relations of the Bessel and Hankel functions, 
$2 \partial_x Z_{n}(x)=Z_{n+1}+Z_{n-1}$ and $(2n/x) Z_{n}(x)=Z_{n-1}-Z_{n+1}$, where $Z_n$ is $J_n$ or $H_n$.

Comparing eqs \ref{eq:C_sca} and \ref{eq:campocev-1} gives the relation between $E_0$ and $C_{sca}$.
\begin{equation}
\left|\vec{E}_{0n}\right|^{2}=
\frac{k}{4\varepsilon^{2}}C_{sca}
\left|
\vec{M}_{n}
\right|^{2}
\label{eq:campocev-1-2}
\end{equation}
Note that, as the fields in different regions are connected due to the continuity condition, when the fields in the region $0$ goes to infinity the same should happen in every region. 
Then, the lasing thresholds can be found as zeros of the determinant of eq \ref{eq:Matrix}, divergences of $C_{sca}$, or divergences of the electric/magnetic fields in an arbitrary position. 
In principle, the last two strategies could be cumbersome because of dark modes or the presence of nodes in the harmonics. However, dark modes can be problematics only for modes with $A_{0n}$ strictly 0. 
In our case, even modes with values of $A_{0n}$ being really small, which can safely be marked as dark modes as $A_{0n}$ is orders of magnitude smaller than the average value per mode, show divergences of $C_{sca}$ easily found numerically. For example, in the upper panel of Figure \ref{fig:9} one can check that even poles with $A_{0n}$ being almost 0, just over the y axis of the figure, can be found numerically as divergences of $C_{sca}$.

\textbf{Modeling the Optical Active Medium\label{sec:modeling_medium}}
The solution of the scattering problem discussed in the previous sections requires the value of the complex dielectric constant $\varepsilon=\varepsilon'+i\varepsilon''$ of each component for the wavelengths of interest.
The imaginary part of it reflects the amount of absorption or emission of light.
In our treatment and due to the temporary dependence used, $\varepsilon''$ has a positive sign for an active medium and a negative sign for an absorbing medium.
The Clausius Mossotti relation determines the dielectric constant of a mixture of molecules in terms of the number of molecules of each type per unit of volume $N_i$ and with molecular polarizability $\alpha_{i}$
 \begin{equation}
 	\frac{\varepsilon-1}{\varepsilon+2}  =  \sum_{i}\left( 4 \pi \frac{N_{i}\alpha_{i}\left(  \omega\right)}{3}\right)
 \end{equation}
 
Considering a diluted solution of an excited dye (subscript $dye$) dissolved in a non-absorbing host medium (subscript $h$), the first order Taylor expansion in $N_{dye}/N_h$ leads to
 \begin{equation}
\varepsilon  \approx  \varepsilon_{h}+\left ( \frac{2+\varepsilon_{h}}{3}\right )^{2} 4 \pi N_{dye} \alpha_{dye}\left(\omega\right)\label{eq:epsilon_Taylor}
\end{equation}

One can take a simple two-level system for modeling the polarizability of the dye,\cite{novotny2012} which, assuming a complete population inversion of the dye and the sign convention for $\varepsilon''$, results in  \begin{equation}
\alpha\left(\omega\right) =
\frac{
\left(\frac{\left|\mu_{12}\right|^{2}}{ \hbar }\right)
\left(\omega_{0}-\omega+i\frac{\gamma}{2}\right)}{\left(\omega_{0}-\omega\right)^{2}+\left(\frac{\gamma}{2}\right)^{2}\left(1+2\left(\frac{\omega_{R}}{\gamma}\right)^{2}\right)}\label{eq:alpha}
\end{equation}
where $\vec{\mu}_{12}$ is the transition dipole moment between the ground and excited states with an energy difference of $\Delta E =\hbar \omega_{0}$, $\gamma$ is a phenomenological damping term, and
$\omega_{R}=\left | \vec{\mu}_{e \rightarrow g} \cdot \vec{E}   \right|/\hbar$ is the Rabi frequency.
As can be noticed, eq \ref{eq:alpha} takes into account the saturation effects. However, in the present paper, we will not take them into account for the following reasons:
Given that the electromagnetic fields inside the structures analyzed (with a large radius) is not homogeneous, the inclusion of saturation effects will introduce spatial inhomogeneities into the dielectric constant of the active medium. This will break the symmetry of the problem, making the expansion of the fields in eq \ref{eqn:3Cz2} not very useful, as now additional boundary conditions are required (see Boundary conditions section).
In such a case, the present approach does not offer advantages over other methods such as finite elements for example.
As discussed in several references,\cite{arnold2015,passarelli2016} the consequence of not taking into account the saturation effects is that fields go to infinity once optical losses are compensated. However, this divergences can indeed be used to find the lasing conditions, which is the main point of the present work.

Combining eqs \ref{eq:epsilon_Taylor} and  \ref{eq:alpha} (neglecting saturation effects) results in
\begin{equation}
\varepsilon=\varepsilon_{h}+\varepsilon_{l}\frac{(\omega_0-\omega+i\frac{\gamma}{2}) \left (\frac{\gamma}{2}
\right ) }{(\omega_0-\omega)^{2}+(\frac{\gamma}{2})^{2}} \label{eq:nosat},
\end{equation}
where $\varepsilon_{l}$ is the maximum value of the imaginary part of the dielectric constant $\varepsilon ''$
\begin{equation}
\varepsilon_{l}=
\left[\frac{2+\varepsilon_{h}}{3}\right]^{2} 
\left(\frac{8 \pi N_{dye} \left|\mu_{12}\right|^{2} }{ \hbar\gamma}\right) \label{eq:epsilon_l}
\end{equation}
Note that, as we are assuming a complete population inversion of the dye, every molecule contributes to the stimulated emission of radiation, thus $N_{dye}$ is directly its concentration.

One additional approximation that is normally used to describe an optical active medium is the wideband approximation, which comes from taking $|\omega-\omega_{0}| \ll \gamma/2$. In this case eq \ref{eq:nosat} turns into 
\begin{equation}
\varepsilon={\varepsilon_{h} + i\varepsilon_{l}} . \label{eq:wideband}
\end{equation}

It should be mentioned that the wideband approximation does not comply with the Kramers Kronig 
relation\cite{hulst1981,bohren2008absorption,novotny2012} nor does it take into account the spectral line of any particular dye.
However, at a maximum of absorption or emission, that is to say at the resonant wavelength of the dissolved dye, it happens that the contribution to the imaginary part of the refractive index presents a maximum while the real part goes to zero.
Under this condition eq \ref{eq:wideband} becomes exact.
The wideband approximation is useful for the first exploration of the system as it allows one to calculate the optical response of a system independently of the characteristic of the active medium. 
Then, after choosing the poles of $C_{sca}$ to be used, a specific dye can be selected, and a better approach can be used.\cite{passarelli2016}

We used water as the host medium ($\varepsilon_{h}=1.77$) in most of the calculations except in Figure \ref{fig:6} where we also use ethanol ($\varepsilon_{h}=1.85$). The values of $\gamma$ and $\varepsilon_{l}$ for the different dyes were adjusted from their emission spectra using the F\"uchtbauer Ladenburg method, as described in the next section.

\textbf{Parameters of the Dyes.\label{sec:dyes}}
While the spectral shape of fluorescence light is relatively easy to measure, it is much more challenging to measure its absolute values due to the difficulty of measuring accurately some quantities such as doping concentration, the degree of electronic excitation, collection and detection efficiencies, etc.
The F\"uchtbauer Ladenburg method\cite{FLmethod} is used to obtain the absolute scaling of the emission cross-section spectrum.
In this method, one exploits the fact that the excited-state lifetime is close to the radiative lifetime $\tau_{rad}$, which itself is determined by the emission cross section $\sigma(\omega)$ for transitions to any lower-lying energy level. This is quantitatively described by the equation
\[
\frac{1}{\tau_{rad}}\approx\frac{2 \varepsilon_{h}}{\pi c^{2}}\int \omega^{2}\sigma(\omega)d\omega .
\]
Knowing the spectral intensity of fluorescence $I(\omega)$, proportional to $\sigma(\omega)$, and using the above equation, one finds
\begin{equation}
\sigma(\omega)=\frac{\pi c^{2}}{2 \varepsilon_{h}\tau_{rad}}\frac{I(\omega)}{\int{\omega}^{2} I(\omega)d\omega} \label{eq:sig_I}
\end{equation}
Typically, the spectra of the dyes $I(\omega)$ are fitted to a Lorentzian function
\[
I(\omega)=\frac{\left ( \gamma/2 \right )^2 }
{  (\omega_{0}-\omega)^{2}+\left ( \gamma/2 \right )^{2} } .
\]
Since, typically, the width of the emission spectrum $\gamma$ is much smaller than the position of the resonance $\omega_0$, one can take the $\omega^2$ factor out of the integrand of eq \ref{eq:sig_I}. This yields
\begin{equation}
\sigma(\omega)=\frac{c^{2}}{\varepsilon_{h}\tau_{rad} \omega_0^{2}\gamma} I(\omega). \label{eq:sig_max1}
\end{equation}

%\begin{widetext}
%	 \centering
\begin{table*}
	\label{table:dyes}
	 	 \caption{Parameters of Dyes Used in Figure \ref{fig:3}.}
 \begin{tabular}{ c c c c c c}
	&    $\quad\bm{\tau_{rad}(ns)}\quad$    & \multicolumn{1}{c}{$\quad\quad\bm{\nu_0}\quad\quad$} & \multicolumn{1}{c}{$\quad\quad\bm{\gamma/2\pi}\quad\quad$  } & \multicolumn{1}{c}{   $\quad\bm{|\mu_{12}|} \bm{(D)}\quad$   } & \multicolumn{1}{c}{   $\quad\bm{N_i}$ $\bm{(\mu M)}\quad$   } \\ \hline \hline	
	{\textit{Alexa-fluor-680}} & 1.200 & 4.27 (702) & 0.226 (13265) & 6.847 & 6.781 \\ \hline
	{\textit{Oregon-green-488}} & 4.100 & 6.03 (497) & 0.331 (9057) & 2.203 & 0.480 \\ \hline
	{\textit{Atto-rho3b}} & 1.500 & 5.03 (596) & 0.336 (8923) & 4.786 & 2.231 \\ \hline
	{\textit{Atto-725}} & 0.500 & 3.96 (757) & 0.266 (11270) & 11.843 & 17.262 \\ \hline
	{\textit{cy3b}} & 2.800 & 5.37 (558) & 0.375 (7995) & 3.171 & 0.878 \\ \hline
	{\textit{tCO}} & 3.000 & 6.40 (468) & 1.248 (2402) & 2.358 & 0.146 \\ \hline
	\end{tabular}	\\
\begin{flushleft}
$N_i$ is the concentration of molecules in the excited state required to obtain $\varepsilon_l=0.05$. The values of $\tau_{rad}$ and the spectra of the dyes, from which we obtained $\nu_0$ and $\gamma$, were taken from ref \cite{spectra}. The values of $\mu_{12}$ are calculated from eq \ref{eq:Mu}. Values in the third and fourth columns are in $10^{14}$ Hz or in nanometers between parentheses.	
\end{flushleft}	
\end{table*}
%\end{widetext}

\begin{figure}[h!]
\begin{centering}
\subfigure{\includegraphics[scale=.9]{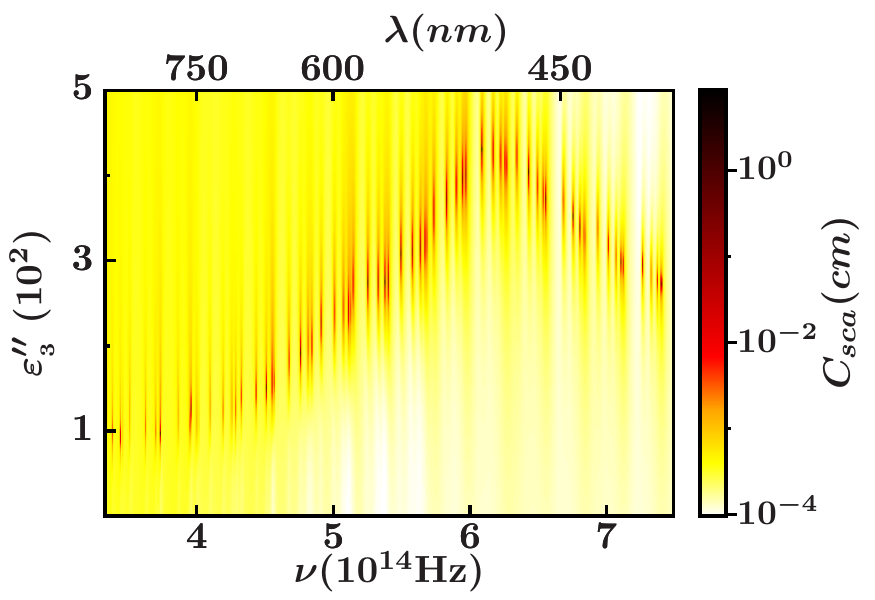}}
\subfigure{\includegraphics[scale=.9]{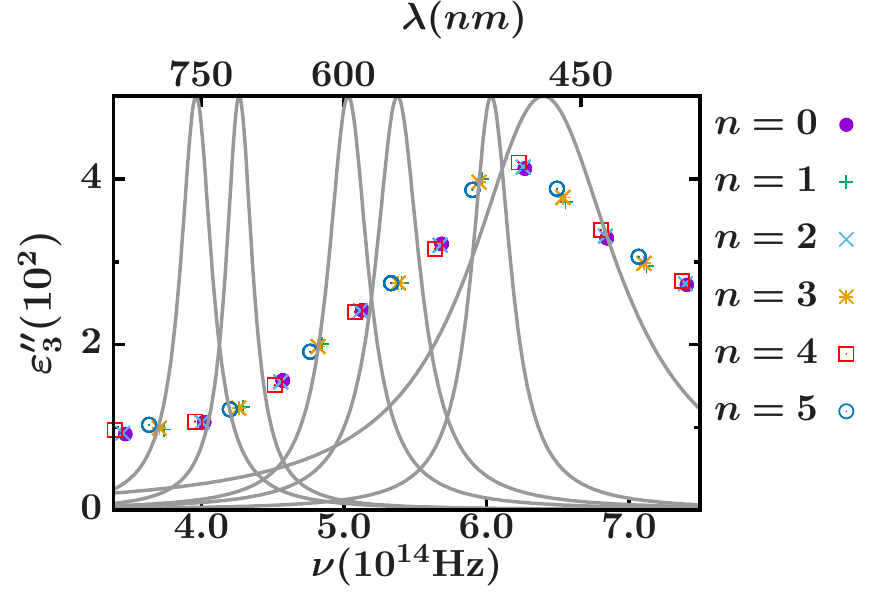}}
\end{centering}
\caption{\textbf{Upper panel: } Scattering cross section $C_{sca}$ as a function of the imaginary part of the dielectric constant of the gain medium $\varepsilon_3''$ and the frequency of the incident light $\nu$.\textbf{ Lower panel:} Position of the poles of $C_{sca}$, for $n$ ($-n$) up to 5 only, obtained within the wideband approximation. Gray lines shows the value of $\varepsilon_3''$ resulting from using different dyes, with $\varepsilon_l=0.05$. From left to right, "Alexa-fluor-680", "Oregon-green-488", "Atto-rho3b", "Atto-725", "cy3b", and "tCO". See the Parameters of the dyes section.}
\label{fig:3}
\end{figure}

On the other hand, the relation between the emission cross section and the polarizability (in cgs) can be written as\cite{novotny2012}
\begin{eqnarray}
\sigma_{aligned}(\omega) & = & \frac{4\pi \varepsilon_h \omega}{c}\mathrm{Im}[ \alpha (\omega) ]. \label{eq:sig_alig}
\end{eqnarray}
where one is assuming a perfect alignment between the electric field and molecule. The maximum value of the macroscopic, orientation averaged cross section $\sigma_{max}$ can be obtained by inserting eq \ref{eq:alpha} into \ref{eq:sig_alig} with $\omega = \omega_0$
\begin{equation}
\sigma_{max} = \frac{\sigma_{aligned}(\omega_0)}{3} 
= \frac{8 \pi \left|\mu_{12}\right|^{2} \varepsilon_h \omega_0}{3 \hbar \gamma c}. \label{eq:sig_max2}
\end{equation}
Comparing eqs. \ref{eq:sig_max1} and \ref{eq:sig_max2} allow us to obtain the value of $\mu_{12}$ that fits the experiments
\begin{equation}
\left | \mu_{12} \right |=\left ( \frac{c}{\omega_0}\right )^{3/2} \left ( \frac{3 \hbar}{8 \pi \varepsilon_h^2 \tau_{rad}} \right )^{1/2} . \label{eq:Mu}
\end{equation}
The dyes used in this work (see Figure \ref{fig:3}) were chosen because they are soluble and present high quantum yields, and their fluorescence spectra and radiative lifetimes are easily accessible. \cite{spectra}
Table \ref{table:dyes} shows the dyes used, the values of $\tau_{rad}$ reported by the manufacturer, the values of $\nu_0$ and $\gamma$ obtained from the fitting of the reported spectra, the values of $\mu_{12}$ obtained from eq \ref{eq:Mu}, and the concentration required to reach $\varepsilon_l=0.05$. The value of $\varepsilon_l$ used ensures the lasing condition over all the frequency ranges studied. Note that the calculation of the concentrations of the dyes shown in the table, of the order of $\mu M$, assumes a complete population inversion, which may require a very intense pumping laser. However, even if only 1 of 1000 molecules is in the excited state the concentrations of the dyes required to ensure the lasing conditions are of the order of $mM$, which does not seem too high considering that all dyes are highly soluble in water.

\begin{figure}[ht]
\begin{centering}
\subfigure{
\includegraphics[width=1.6 in,trim= 0.0in 0.0in 0.0in 0.0in,clip]{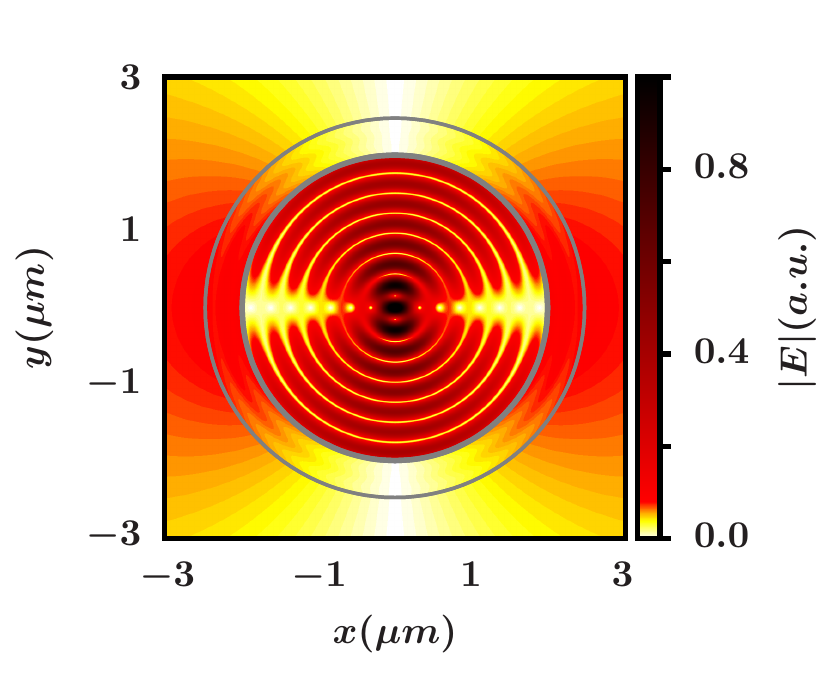}
\includegraphics[width=1.6 in,trim= 0.0in 0.0in 0.0in 0.0in,clip]{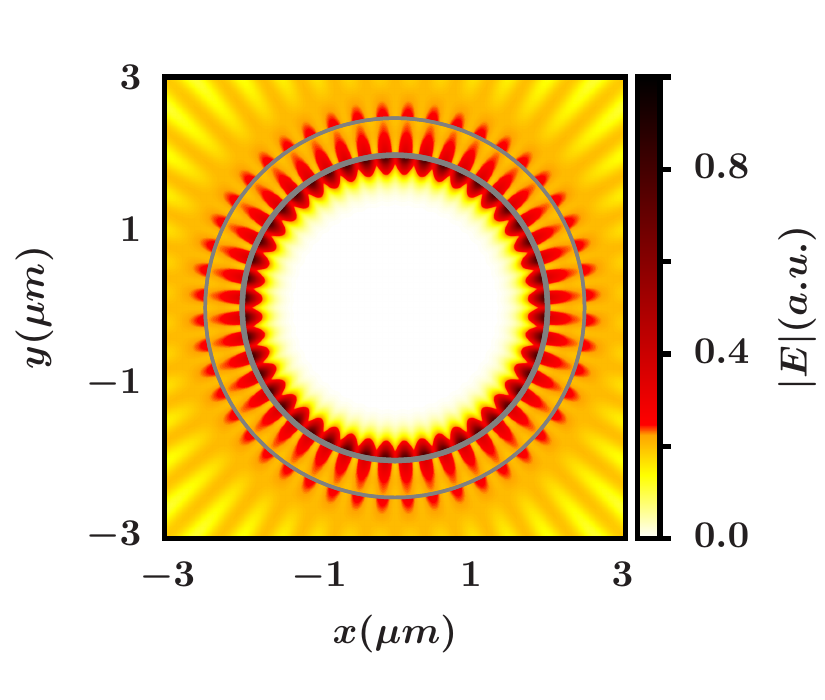}
}
\end{centering}
\caption{
\textbf{Left panel:} Near field of a typical cavity mode close to its gain threshold, $\nu = 4.29 \times 10^{14}\mathrm{Hz}$ $\varepsilon_3'' =0.0124$. \textbf{Right panel:} Near field of a typical whispering-gallery mode close to its gain threshold, $\nu = 4.25 \times 10^{14}\mathrm{Hz} $ $\varepsilon_3'' =0.1532$. The geometry is that of Figure \ref{fig:1} with $R_3 = 2000 \mathrm{nm}$, $R_2-R_3 = 25 \mathrm{nm}$, and $R_1-R_2 = 475 \mathrm{nm}$. The electric fields are normalized to their respective maximum values.
}
\label{fig:2}
\end{figure}

\textbf{Lasing Conditions\label{sec:lasing_conditions}}
Within the interval of parameters explored (see the Results and Discussion section), the analyzed system presents two types of modes: the cavity and the whispering-gallery modes (CMs and WGMs, respectively).
The former corresponds to modes where the fields are mostly concentrated within the region of the active media, and there is a wave pattern along the radius.
The WGMs are modes where the fields are concentrated close to the metal layer, and there is a wave pattern along the circumference of the cylinder.
Typical near fields of both types of modes are shown in Figure \ref{fig:2}.
Note that the lasing of WGMs turns the system into a spaser,  where the light is highly confined close to the metal layer due to the excitation of surface plasmon polaritons.
On the other hand, and due to the fields distribution, the lasing of CMs turns the system into what can be considered a metallic-coated nanolaser (or microlaser depending on the sizes).\cite{ma2018}
Therefore, the proposed system has the peculiarity of being, at the same time, a spaser and metallic-coated nanolaser, depending on which mode is lasing.

As we will see in the Results and Discussions section, the cavity modes present far lower gain thresholds, which, due to the large number of poles, makes the whispering-gallery modes unreachable within the visible spectrum, for the main geometry studied. For that reason, we will focus on the lasing conditions of cavity modes.

\textbf{Lasing Frequencies of Cavity Modes.}
To find the resonant frequencies of the system, we will assume a cavity-like model where the standing wave pattern arises from that of waves confined in a cylindrical cavity.
This is justified by the gold layer surrounding the active medium. Under this model and neglecting the imaginary part of $k_3$, the resonant condition occurs when the argument of the Bessel function $J_n$ corresponding to region 3 ($k_3 r$) becomes zero at the boundary between regions 2 and 3, $r=R_3$, that is
\begin{equation}
  \omega_{n,i} = \frac{c}{m_h R_3} u_{n,i} \label{eq:BZ}
\end{equation}
where $u_{n,i}$ is the $i$-th non trivial zero of the $J_n$ Bessel's function, we are taking $m_3 \approx m_h$ ($m_h$ is the refractive index of the host of the active medium), $R_3$ is defined in Figure \ref{fig:1}, and $\omega_{n,i}$ is the resonant angular frequency. We will call this approximation ``Bessel's zero approximation'' (BZA).
When the radius of the cylinder is large with respect to the wavelength, one can use the asymptotic 
forms of the Bessel functions to describe the radial component of the field of each mode.
\begin{equation}
J_n \left ( \frac{ \omega}{c} m_h r \right ) \propto \cos \left ( \frac{ \omega}{c} m_h r - n \frac{\pi}{2}-\frac{\pi}{4} \right ) \label{eq:BZ_cos}
\end{equation}
Under this approximation, the resonant frequencies result simply
\begin{equation}
 \omega_{n,p} =  \left ( \frac{ c \pi }{4 m_h R_3}  \right )  \left ( 2 p + 2 n +1 \right ) \label{eq:FP}
\end{equation} 
where $p$ is an odd integer. We will call this Fabry P\'erot approximation  as it coincide with the resonances of a Fabry-P\'erot interferometer.

In eqs \ref{eq:BZ} and \ref{eq:FP}, we assumed that the fields at the boundary of regions 2 and 3 were strictly zero and neglected the imaginary part of $k_3$, which is not the case for realistic systems. To correct the formulas, we propose a minimum model that only adds a constant shift $\delta \nu_{corr}$ to the lasing frequencies
\begin{equation}
\nu =\frac{c u_{n,i} }{2 \pi m_h R_3} + \delta \nu_{corr} \label{eq:deltanu}.
\end{equation}
Note, that, because of consistency, the same correction should be applied to both eqs \ref{eq:BZ} and \ref{eq:FP}.
As we will see, introducing $\delta \nu_{corr}$ improves considerably the agreement with numerical results, but in exchange, now one should know \textit{a priori} at least one lasing frequency. Despite this, one is characterizing the whole spectrum of resonant frequencies by a single parameter.

Withing the Fabry P\'erot approximation, the proportionality between $\omega$ and $p$ causes the different modes to be equispaced in frequency. Considering the geometry studied, $m_3 \approx m_{h}=1.33$ and $R_3=2.0 \mu m$: the gap between two contiguous zeros ($p$ and $p+2$) belonging to the same mode $n$ is
\[
\nu_{p+2}-\nu_{p}=4 \times \frac{c}{(8m_{h}R_3)} \simeq 0.56\times 10^{14} \mathrm{Hz}  .
\]
This means that there are approximately seven poles for each $n$ mode within the visible spectrum, for $n$ small. This large amount of poles shows that, given a dye, there will always be some resonance close to the maximum of its emission cross section.

Given the sizes of the tubes used in the calculations, where the largest one has a radius of $2000$ nm, and the wavelengths considered, between $400$ and $800$ nm, deviations with respect to the Fabry P\'erot approximation are expected. We will analyze this in more detail in the Results and discussion section.

\textbf{Gain Thresholds of Cavity Modes.}
To find an expression for the gain threshold we will use the same cavity-like model as before.
We start by the gain loss compensation condition
\begin{equation}
W_{abs}=-W_{sca}
\end{equation}
where $W_{abs}$ and $W_{sca}$ are the energy absorbed and scattered per unit time. The former can be calculated from
\begin{eqnarray}
W_{abs} & = & -\frac{c}{8\pi} \varepsilon''_{3} \int_{V \in \ 3}k \left|\vec{E}\right|^{2}dV \notag \\
&   & -\frac{c}{8\pi} \varepsilon''_{2} \int_{V \in \ 2}k \left|\vec{E}\right|^{2}dV \label{eq:Wabs}
\end{eqnarray}
\newline
\ 
\newline

where indexes $2$ and $3$ refer to regions $2$ an $3$ in Figure \ref{fig:1}, respectively.
The energy scattered per unit time $W_{sca}$ can be calculated from
\begin{eqnarray}
W_{sca} & = &\frac{c}{8\pi}\frac{4lk_{0}^{2}}{k}\sum_{n=-\infty}^{\infty} \left|A_{0n}\right|^{2} \notag \\
& \approx & \frac{c}{\pi}l\varepsilon_0k\left|A_{0n}\right|^{2} 
\times
\begin{cases}
 1 & n \neq 0 \\
 \frac{1}{2} & n = 0
\end{cases}  \label{eq:Wsca}
\end{eqnarray}
where we are using the fact that, at the lasing condition, two coefficients of the summation dominate: $n$ and $-n$.
Using eqs \ref{eq:Wabs} and \ref{eq:Wsca}, one can write the required condition for $\varepsilon''_3$ to reach the gain threshold. For $n \ne 0$ it results in

\begin{equation}
\varepsilon''_{3}=\frac{ 8\varepsilon_0 - \varepsilon''_{2} \int_{V \in 2} \frac{\left|  \vec{E}_n \right |^2 }{l \left|A_{0n}\right|^{2} } dV } { \int_{V\in3}\frac{\left|\vec{E}_{n}\right|^{2}}{l\left|A_{0n}\right|^{2}}dV }  , \label{eq:epsilon_3}
\end{equation}

where $E_n$ is the electric field produced by the excitations of modes $n$ and $-n$. For $n = 0$ the factor $8$ should be replaced by a factor $4$.

The integrals over the electric field in eq \ref{eq:epsilon_3} can be written in terms of the Bessel and Hankel functions. The result for $n \neq 0 $ is

\begin{widetext}
\begin{eqnarray}
\int_{V\in2}\frac{\left|\vec{E}_{n}\right|^{2}}{l\left|A_{0n}\right|^{2}}dV&=&
2\pi \intop_{R_{3}}^{R_{2}} \left [
\frac{\left|a_{2n}\right|^{2}}{\left|A_{0n}\right|^{2}}\left(\left|J_{n-1}\left(k_{2}r\right)\right|^{2}+\left|J_{n+1}\left(k_{2}r\right)\right|^{2}\right) 
\right .
\notag  \\
&&+\mathrm{Re}\left\{ \frac{ A_{2n}
a_{2n}^{*}}{\left|A_{0n}\right|^{2}}\left[H_{n-1}\left(k_{2}r\right)J_{n-1}^{*}\left(k_{2}r\right)+H_{n+1}\left(k_{2}r\right)J_{n+1}^{*}\left(k_{2}r\right)\right]\right\}
\notag  \\
&&
\left .
+\frac{\left|A_{2n}\right|^{2}}{\left|A_{0n}\right|^{2}}\left(\left|H_{n-1}\left(k_{2}r\right)\right|^{2}+\left|H_{n+1}\left(k_{2}r\right)\right|^{2}\right)
\right ]
\left|k_{2}\right|^{2}rdr
\label{eq_EV2}
\end{eqnarray}
and
\begin{equation}
\int_{V\in3}\frac{\left|\vec{E}_{n}\right|^{2}}{l\left|A_{0n}\right|^{2}}dV
=
2\pi\frac{\left|a_{3n}\right|^{2}}{\left|A_{0n}\right|^{2}}\intop_{0}^{R_{3}}\left(\left|J_{n-1}(k_{3}r)\right|^{2}+\left|J_{n+1}(k_{3}r)\right|^{2}\right)\left|k_{3}\right|^{2}rdr
\label{eq_EV3}
\end{equation}
\end{widetext}

where we have used the recurrent relations of $J_n$ and $H_n$. 
For $n = 0$ a $1/2$ factor should be added to the right-hand side of both equations.

As we are dealing with cavity modes where the electric field is concentrated mostly within the cavity, one can assume that the integral of $|E|^2$ over region 2 in eq \ref{eq:epsilon_3} is negligible. Then, eq \ref{eq:epsilon_3} turns into 
\begin{equation}
\varepsilon''_{3} \approx
\frac{
\left (
\frac{4\varepsilon_0 \left|A_{0n}\right|^{2}}
{\pi \left|a_{3n}\right|^{2}}
\right )
}
{
\intop_{0}^{R_3} \left(\left|J_{n-1}\right|^{2}+\left|J_{n+1} \right|^{2}\right) |k_3|^2 r dr
}. \label{eq:epsilon_3-2}
\end{equation}

\begin{figure}[h!]
\begin{centering}
\subfigure{\includegraphics[scale=.9]{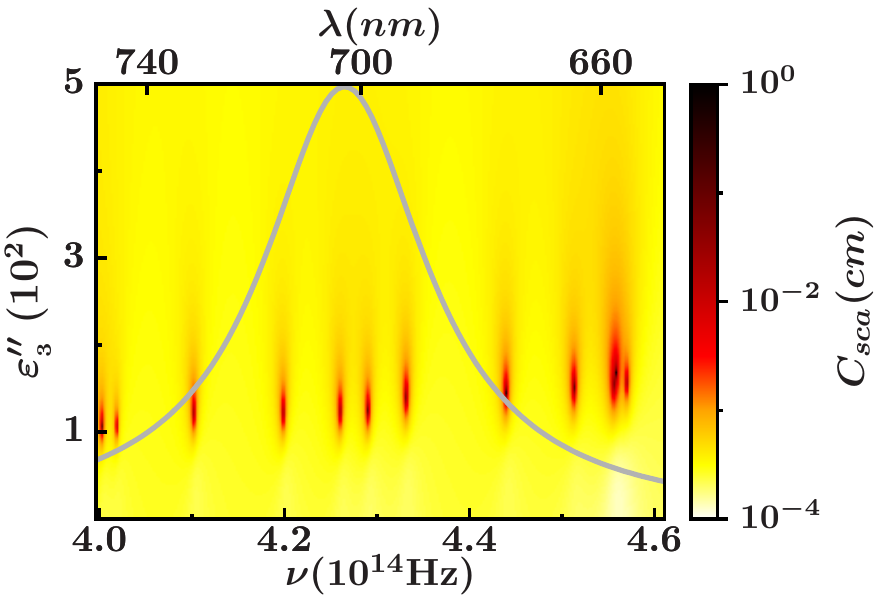}}
\subfigure{\includegraphics[scale=.9]{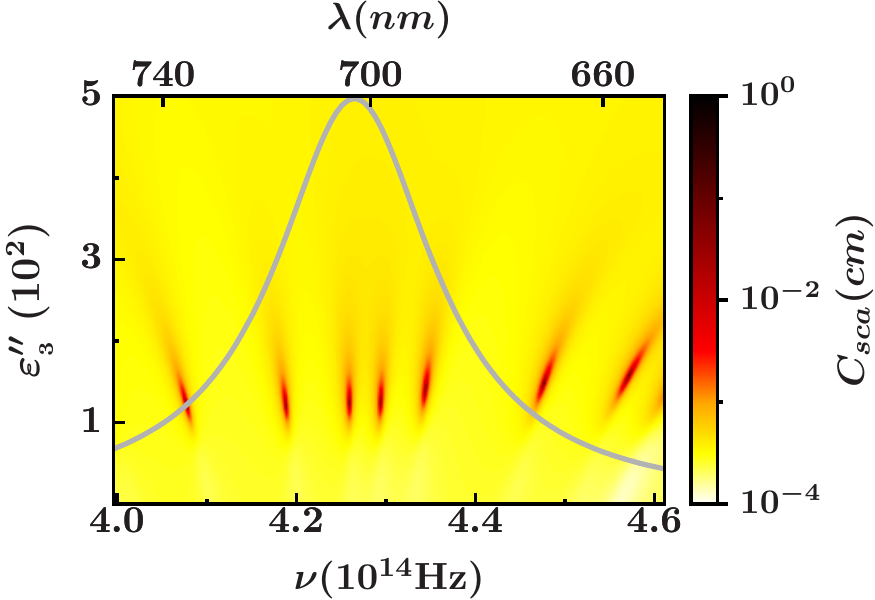}}
\end{centering}
\caption{\textbf{Upper panel:} $C_{sca}$ as function of $\varepsilon_3''$ and $\nu$ obtained using the wideband approximation (eq \ref{eq:wideband}).
\textbf{Lower panel:} Same as the upper panel but $C_{sca}$ was calculated using eq \ref{eq:nosat} with Alexa-fluor-680 as the dye.
Gray lines in both figures are the values of $\varepsilon_3''$ obtained using Alexa-fluor-680, with $\varepsilon_l=0.05$.
}
\label{fig:4}
\end{figure}

Note that the right hand side of eqs \ref{eq:epsilon_3} and \ref{eq:epsilon_3-2} depends on $\varepsilon''_{3}$, as the imaginary part of $k_3$ does. Therefore, the equation have to be solved either iteratively or by an optimization method. One approximation that can be made to avoid that is to use the BZA discussed in the previous section: that is, one can use eq \ref{eq:deltanu} to estimate $k_3 \in \mathrm{Re}$. Given a value of $k_3$, and thus $k_0$, one can solve the set of linear equations given in eqs \ref{eqn:3CFconti1}-\ref{eqn:3CFconti6} to obtain the $a_{jn}$ and $A_{jn}$ coefficients as well as the integrals involving the Bessel and Hankel functions in eq \ref{eq:epsilon_3} (or eq \ref{eq:epsilon_3-2}).

Despite the different approximations that one can make to estimate \textit{a priori} the value of $\varepsilon_3''$, which, as we will see ,do not work that well, eqs \ref{eq:epsilon_3} and \ref{eq:epsilon_3-2} can be useful for other reasons, for example to find the lasing conditions by directly applying an optimization method on eqs \ref{eq:epsilon_3}. Furthermore, the equations can also help rationalize the observed trends in the gain threshold of different systems. For examples, according to eq \ref{eq:epsilon_3-2}, concentrating the fields on the active medium region (increasing $|a_{3n}|^2$) should reduce the gain thresholds.

\section{Results and Discussion\label{sec:results}}
\begin{figure*}
\begin{centering}
\includegraphics[scale=1.1]{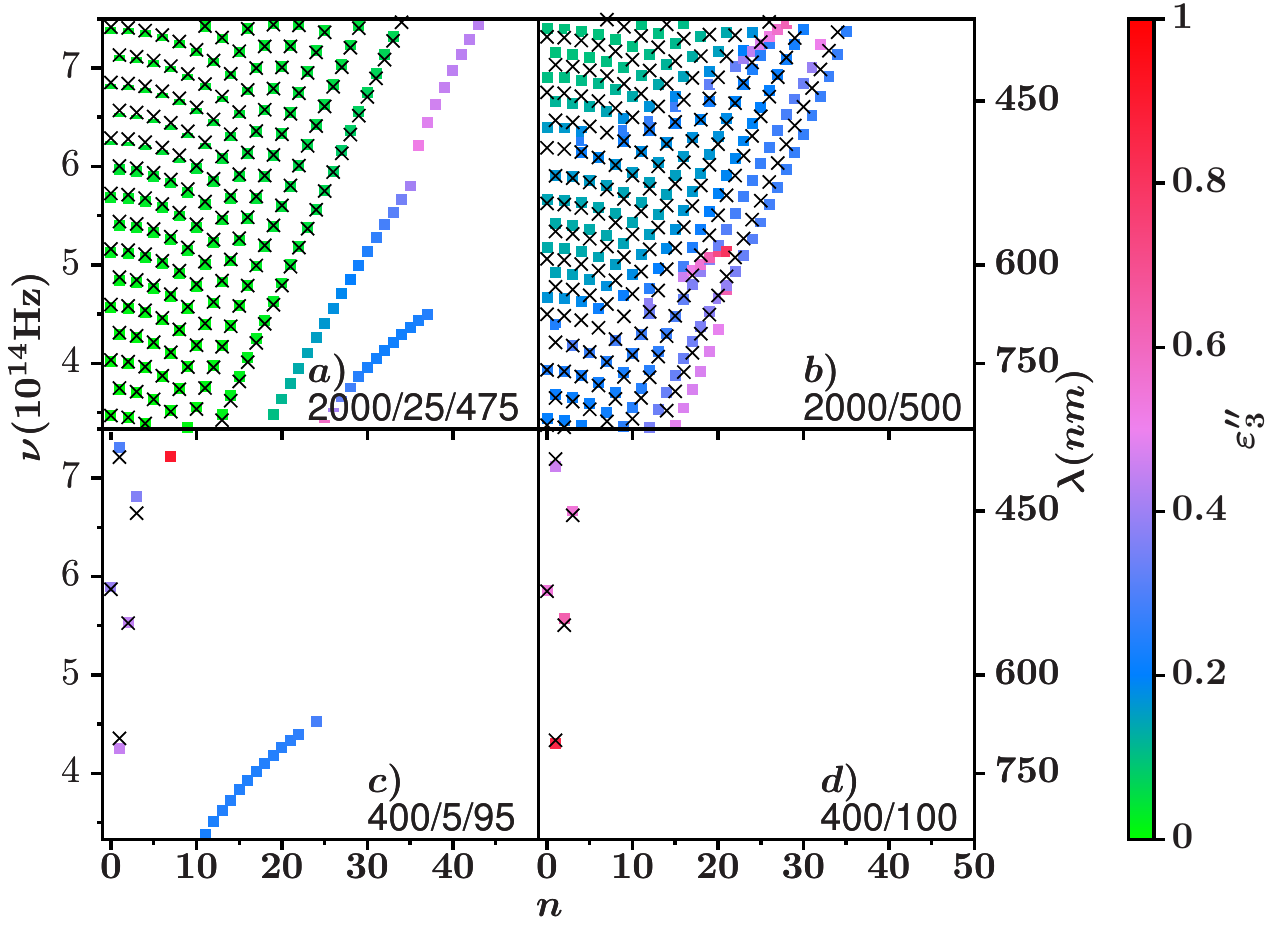}	
\end{centering}
\caption{
Filled dots corresponds to the position of the poles, $\nu$ and $\varepsilon_3''$, found numerically for different geometries and modes $n$. The crosses are the values of $\nu$ obtained using the BZA (eq \ref{eq:BZ}). Geometries are indicated above each figure, for example, panel (a) corresponds to $R_3 = 2000 \mathrm{nm}$, $R_2-R_3 = 25 \mathrm{nm}$, and $R_1-R_2 = 475 \mathrm{nm}$. The values of $\delta\nu_{corr}$ used are $0.23$, $0.13$, $0.92$, and $0.90$ ($10^{14}\mathrm{Hz}$) for (a)-(d) respectively.
In (a) and (c), filled dots without crosses nearby (at large $n$) correspond to poles of whispering-gallery modes.
}
\label{fig:8}
\end{figure*}

\begin{figure}[ht]
\includegraphics[scale=.9]{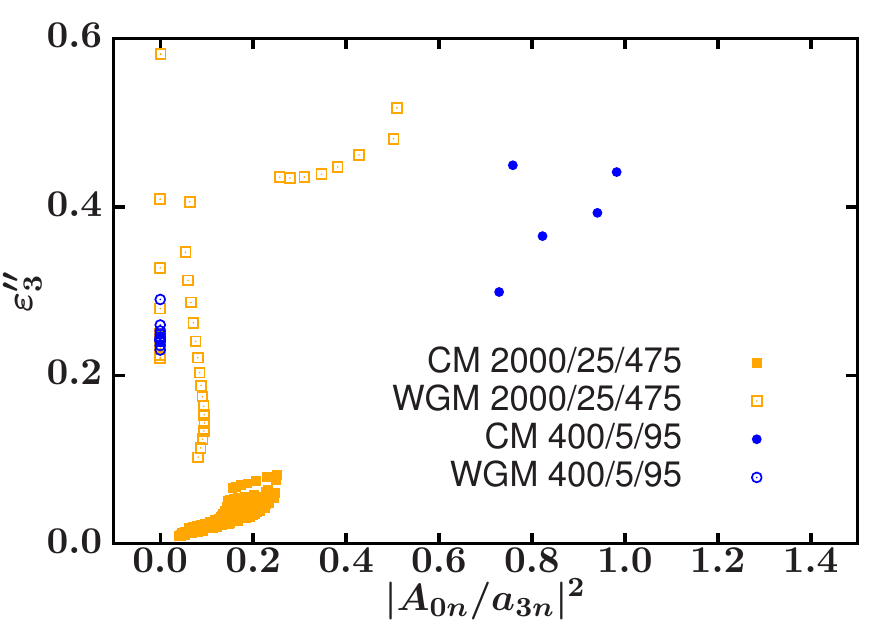}
\includegraphics[scale=.9]{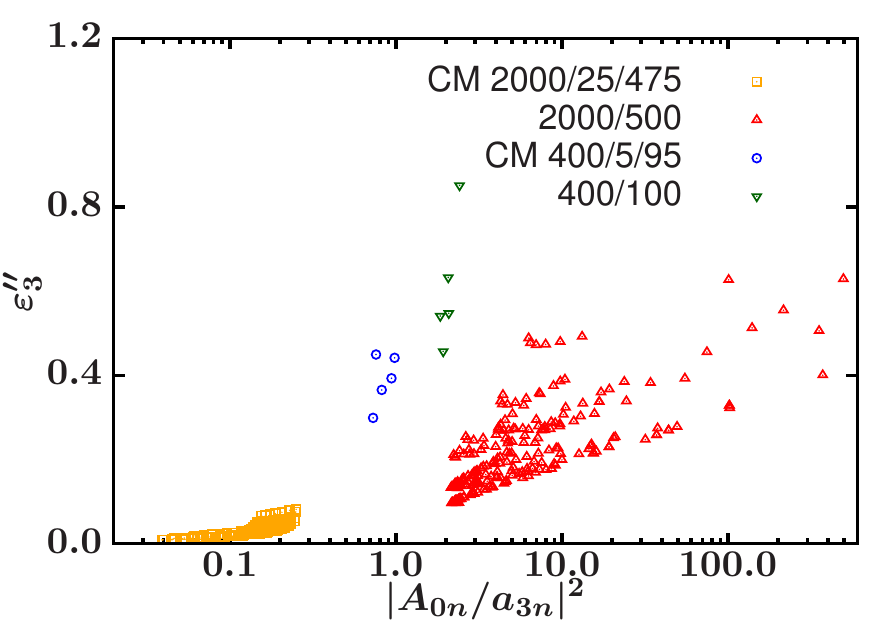}
\caption{
Imaginary part of the dielectric constant of the active medium, $\varepsilon_3''$, as a function of the confinement, calculated as $|A _{0n}|^2/|a_{03}|^2$, for different geometries and types of modes, indicated in the insets.
}
\label{fig:9}
\end{figure}

Using the methodology described in the previous sections, we study different geometries of multiwalled micro- and nanotubes.
In all the figures, except in Figures \ref{fig:8} and \ref{fig:9}, we used, as a representative example, the geometry shown in Figure \ref{fig:1} and discussed in the  General remarks section.  
We first used the wideband approximation(eq \ref{eq:wideband}) to find the poles of the system. The upper panel of Figure \ref{fig:3} shows the scattering cross section $C_{sca}$ as a function of the imaginary part of the dielectric constant of the gain medium $\varepsilon_3''$ and the frequency of the incident light $\nu$. As discussed previously, the poles of $C_{sca}$ correspond to the full loss compensation condition, or the lasing condition.
These conditions are difficult to visualize in this figure due to the large number of poles that the system present.
In the lower panel of Figure \ref{fig:3}, we show only the position of the poles, to better visualize them.
\begin{figure}[h!]
	\includegraphics[scale=.9]{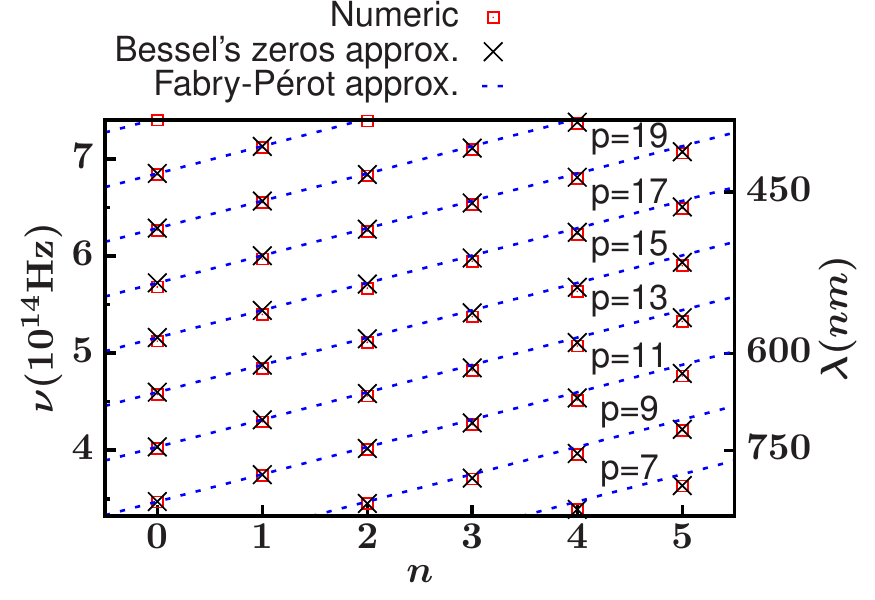}
	\caption{Comparison of the values of $\nu_{pole}$ obtained numerically, with the BZA, (eq \ref{eq:BZ}) and the Fabry P\'erot approximation (eq \ref{eq:FP}). For both  approximations, we used a constant shift of $\delta \nu_{corr}=0.23 \times 10^{14} \mathrm{Hz}$. The values of $p$ used in the Fabry P\'erot approximation are indicated just above the corresponding blue lines. 
	} \label{fig:5}
\end{figure}

To find the position of the poles, we used a simplex\cite{NRecipies} algorithm to maximize the function $C_{sca}(\nu,\varepsilon'')$, starting from the maxima of $C_{sca}$ obtained from a systematic variation of $\nu$ and $\varepsilon_3''$.
To distinguish true divergences from simple maxima, we use the fact that, in a pole, the iterative process will change drastically the value of $C_{sca}$, meanwhile in a maximum the value should converge without much changes in $C_{sca}$. We consider a maximum as a true pole only if $C_{sca}$ changes more than $100$ times its initial value.
We checked the poles found with this procedure and confirm they are in perfect agreement with eq \ref{eq:epsilon_3}.

The upper panel of Figure \ref{fig:3} shows the enormous quantity of poles that can be achieved with relatively low gains, lower than 0.05. In the lower panel of the figure, we only show the poles up to order 5 for clarity, but in total there are around 200 poles within the visible spectrum.
Superimposed on the figure, we also show the imaginary part of the dielectric constant of the active media resulting from using different dyes, see the Parameters of the dyes section.
If we consider that the dyes only contribute to the imaginary part of the dielectric constant of the gain medium, all the poles below a gray curve will be reached with the dye and concentration being studied.
As can be seen, it is possible to obtain TE modes of lasing all along the visible spectrum, and using moderate concentrations of the dyes, see Table \ref{table:dyes} in section ``Parameters of the dyes''.
\begin{figure}[h!]
\includegraphics[scale=.9]{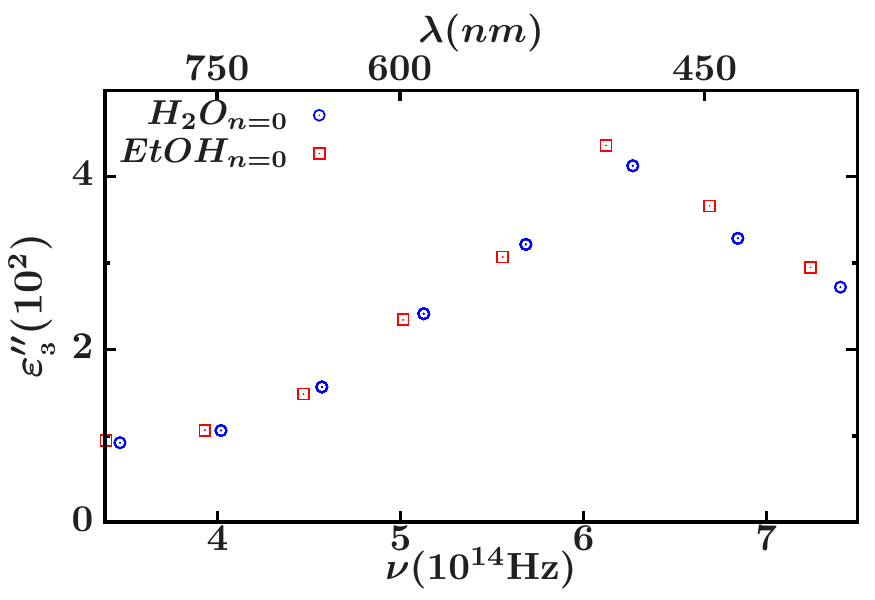}
\caption{Example of the variation of the position of the poles induced by a change in $\varepsilon_h$.}	\label{fig:6}
\end{figure}

It is interesting to compare the position of the poles of $C_{sca}$ obtained with the wideband approximation with that obtained with the explicit consideration of a particular dye using eq \ref{eq:nosat}. This is done in Figure \ref{fig:4}. As expected, when the resonant frequency of a pole $\nu_{pole}$ is close to the maximum of the emission cross section of the dye, the results for both approximations coincide. However, even when $\nu_{pole}$ is quite far from the maximum, the wideband approximation seems to work pretty well especially at predicting the values of $\varepsilon''_{pole}$. Only small deviations in the values of $\nu_{pole}$ were observed in such cases. This is reasonable once one realizes that, due to the low concentration of the dyes, the real part of the dielectric constant is basically that of the host medium $\mathrm{Re}\left ( \varepsilon \right ) \approx \varepsilon_h$, while its imaginary part is small, around 3\% of $\left | \varepsilon \right |$.
\begin{figure}[h!]
\begin{centering}
\includegraphics[scale=.9]{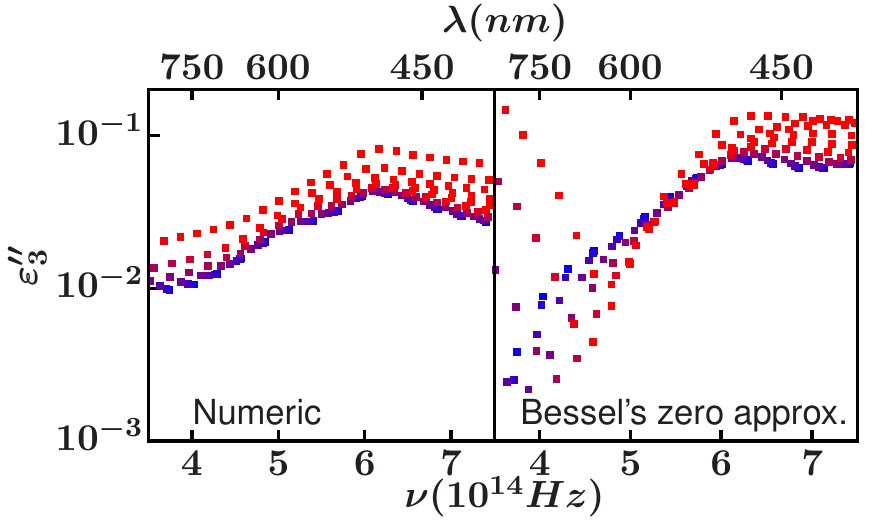}
\end{centering}
\caption{
\textbf{Left panel:} Position of the poles of $C_{sca}$ for all $n$ ($-n$) obtained numerically. \textbf{Right panel:} Same, but poles are obtained using Bessel's zeros approximation, that is $\nu$ is calculated from eq \ref{eq:deltanu} ($\delta \nu_{corr} = 0.23\times10^{14}\mathrm{Hz}$), and $\varepsilon_3''$ is calculated from eq \ref{eq:epsilon_3-2} with $k_3 \in \mathrm{Re}$ estimated from eq \ref{eq:deltanu}. The color of the dots stands for the order $n$, and low $n$ are plotted in blue and high $n$ in red.
}
\label{fig:7}
\end{figure}

Figure \ref{fig:5} compares the position of the poles obtained numerically, maximizing $C_{sca}(\nu,\varepsilon'')$, with the positions predicted by the BZA (Eq. \ref{eq:BZ}) and the Fabry P\'erot approximation (eq \ref{eq:FP}). As can be seen, the agreement with the BZA is almost perfect.

On the other hand, the agreement with the Fabry P\'erot approximation is not bad at low $n$. However, at higher $n$, the agreement worsens, and modes with the same value of $p$ significantly deviate from the predicted linear dependence with respect to $n$. Despite this, this simple approximation estimate, surprisingly well, the frequency separation between modes with the same $n$ but different $p$, even at high values of $n$.

Note that in Figure \ref{fig:5}, only one parameter has been used to adjust both approximations: the global correction to the frequency of the poles $\delta \nu_{corr}$.

Besides its obvious application for enhanced spectroscopies, the type of system proposed can be used for sensing, through changes in the dielectric constants.\cite{kedem2014,kim2013cavity,spackova2016} In this case, the fact that the width of the peaks in the spectrum of micro- and nanolasers are very narrow is a great advantage for the sensitivity of any techniques based on the shift of peaks. Figure \ref{fig:6} shows the large shift of the poles with $n=0$ caused by a slight change in the dielectric constant of the host medium $\varepsilon_h$, from 1.77 (water) to 1.85 (ethanol).

From the values of Fig. \ref{fig:6}, it is possible to estimate the ``refractive index sensitivity'' (RIS), widely used for sensor characterization and defined as the ratio between the resonance shift $\Delta \lambda$ and the refractive index variation of the analyte, $\Delta n$.\cite{kedem2014,spackova2016} RIS is usually measured in nm/RIU, where RIU is the abbreviation of ``refractive index unit''. In Figure \ref{fig:6}, the RIS ranges from $305$ to $648$ nm/RIU for $n=0$, which is a relatively large value according to the bibliography.\cite{kedem2014}
In principle eq \ref{eq:FP} could have been used to predict the variation of the position of the peaks, and the values of RIS.
However, this does not lead to good estimations, since the global shift used to correct eqs \ref{eq:BZ} and \ref{eq:FP}, $\delta \nu_{corr}$, depends not only on the geometry of the system but also on its dielectric constants. For example, according to eqs \ref{eq:BZ} the change in the position of the peak at the higher frequency should be $0.23\times 10^{14}\mathrm{Hz}$, but the actual value is $0.162\times 10^{14}\mathrm{Hz}$.

Figure \ref{fig:7} compares the lasing conditions obtained numerically with those found using the BZA for to the calculation of both, $\nu$ (eq \ref{eq:deltanu} with $\delta \nu_{corr}= 0.23 \times 10^{14} \mathrm{Hz}$) and $\varepsilon''_3$ (eq \ref{eq:epsilon_3-2} with $k_3$ given by eq \ref{eq:deltanu}). As can be seen, the predicted values of $\varepsilon''_3$
deviate significantly for the lowest frequency modes where there is an unrealistic widespread of $\varepsilon''_3$ values. For intermediate and high frequencies (larger than 5 Hz approximately) the differences are smaller and at least the BZA estimates well the order of magnitude of $\varepsilon''_3$. The reason for the large errors is that both, the coefficients $a_{jn}$ and $A_{jn}$ and the integrals over the Bessel and Hankel functions, are very sensitive to the exact complex value of $k_3$.

To compare the effect of the geometry of the system on the lasing conditions, in Figure \ref{fig:8} we show the position of the poles corresponding to four different geometries. The geometry ``2000/500'' is the same as ``2000/25/475''(the geometry studied so far) but the gold layer has been replaced by a thicker polymer layer. The geometry ``400/5/95'' is the same as ``2000/25/475'' but rescaled by a factor 5. The same is true for the last geometry, ``400/100'', with respect to ``2000/500''. In Figure \ref{fig:8}, the crosses indicated the values of $\nu$ predicted by the BZA while the squares are the values of $\nu$ found numerically. In Figure \ref{fig:8} a, c, the squares without crosses nearby correspond to WGMs. As shown in the figure, the BZA is able to predict the resonant frequency $\nu$ of all the cavity modes but only when there is a metal layer. For systems without a metal layer (see Figure \ref{fig:8} b, d) the BZA can also work but only in limited spectral regions.
As mentioned at the beginning of the Lasing conditions section,
the WGMs have much higher gain thresholds  than those of cavity modes for the main geometry studied(Figure \ref{fig:8}a). However, for smaller systems, the increase of the gain thresholds of cavity modes makes them comparable. This is better appreciated in the upper panel of Figure \ref{fig:9} where one can check that the values of $\varepsilon_3''$ of cavity modes and WGMs are comparable for the geometry 400/5/95.

Figure \ref{fig:8} can also be used to analyze the effects of different geometries on the gain threshold of cavity modes. Broadly speaking, the figure shows that the gold layer reduces the gain thresholds, while reducing the size of the system increases the gain thresholds.
This is better appreciated in Figure \ref{fig:9} where we plot $\varepsilon''_3$ as a function of $|A_{0n}|^2/|a_{3n}|^2$, which can be taken as an indicator of the confinement of the fields.
Although a rigorous comparison of gain thresholds is difficult for different geometries, since the resonant frequency of the same modes changes with the geometry, the general trend for cavity modes (CM in the figure) is ``the more confined the field, the lower the gain threshold'', or to be more precise, smaller values of $|A_{0n}|^2/|a_{3n}|^2$ are correlated with smaller values of $\varepsilon''_3$.
This is in agreement with the expected trend of eq \ref{eq:epsilon_3-2}, although the lack of a proportionality relation indicates that the numerator of eq \ref{eq:epsilon_3-2} varies significantly with the mode and the geometry of the system, or that the power dissipated by the gold layer is not completely negligible. By inspection of our results, we concluded that both contributions are responsible for the deviations. For WGM, it does not seem to be a correlation between $|A_{0n}|^2/|a_{3n}|^2$ and $\varepsilon''_3$ (see upper panel of Figure \ref{fig:9}) which is natural considering that the power dissipated by the gold layer plays a central role in determining the gain thresholds in this case.

\section{Conclusions\label{sec:conclusions}}

The studied geometry presents two types of TE modes: cavity modes and whispering-gallery-modes. Within the visible spectrum, there is a large number of both. Cavity modes present very low gain thresholds ($\varepsilon_3''$ around $10^{-2}$) while the WGMs exhibit gain thresholds an order of magnitude higher. 
The small separation between consecutive cavity modes and their much smaller gain thresholds makes WGMs hard to use for lasing.
The reason for that is that, because of the finite width of the emission cross- section of the active medium, the population inversion will be depleted by some cavity mode way before the lasing conditions of a WGM is reached.\cite{passarelli2016}
Therefore, for the type of systems studied here, only in tubes with radii in the nanoscale, the WGMs may be used for lasing.

The very low gain thresholds of cavity modes imply that the required concentration of the dyes can be as low as $10^{-6} M$.
On the other hand, the geometry studied allow one the recycling of the active medium and thus can be used to counteract the effects of photobleaching, an effect that may limit the practical applications of other proposals such as active core shell nanoparticles.

In the geometry studied, the frequencies of lasing can always be tunned by changing the geometry of the system but also by changing the dielectric constant of the gain medium using different solvents for example. The latter can also be useful for sensing purposes. On the other hand, given that the system poses so many tunable resonances, it should not be difficult, in principle, to tune a system where the maximum of the absorption cross section of the dye and the frequency of the pump laser coincide with one resonance, while the maximum of the emission cross section of the dye coincides with another one. This should greatly enhance the performance of the devices.

We have proposed a simple model that can accurately predict the wideband limit of the frequencies of lasing for the geometry studied. This may be especially useful in the design of the devices. On the other hand, we have shown that the deviations from the wideband approximation are not so important even for modes whose frequencies do not coincide with the maximum of the emission cross-section of the dyes.

Comparing different geometries, we have found that increasing the size of the systems, considerably reduces the gain threshold and increases the number of lasing frequencies within a given frequency interval. We have also shown that adding even a very small layer of a plasmonic material such as gold (a 25 nm layer in a microtube whose total radius is 2.5 $\mu$m) dramatically reduces the gain threshold of the nanolaser (up to one order of magnitude).
The effect of the geometry on the gain thresholds can be rationalized, at least qualitatively, in terms of the confinement of electromagnetic fields into the region of the active media.

Contrary to the longitudinal modes of lasing, which only emits plane waves, the electromagnetic fields emitted by the TE modes present hotspots without requiring additional nanostructures. Those hot spots are relatively large and can easily accommodate large macromolecules or even greater analytes such as entire cells.
Moreover, the electromagnetic fields can be further enhanced with the aid of adsorbed nanoparticles, if required.
In micro- and nanolasers, longitudinal modes can, in principle, interfere with the TE modes since they compete for the excited states to produce stimulated emission of radiation. However, their gain threshold can be independently changed by controlling their confinement, for example by leaving or not the tubes opened at the ends or using absorbing materials there.
A possible future research direction is to compare, for the same system, the lasing conditions of longitudinal and transverse modes, addressing explicitly the problem of competition for the excited states.

In this work, we used the divergences of the scattering cross section to find the spasing conditions, but it would be computationally useful to compare this strategy with other methods, for example using point-dipole emitters to look for divergences of near fields, or directly solving iteratively (eq \ref{eq:epsilon_3}) which just express compensation of gains and losses. Moreover, for theoretical completeness, it is important to study the scattering of multiwalled nanotubes for oblique illumination.

Although the system studied seems promising for several applications such as sensing or as part of a ``lab-on-chip'' device, more studies are required to evaluate, for example, the intensity of the electromagnetic fields attainable for the proposed geometry, or the true potential of the system for sensing purposes.

Ultimately it would be important to experimentally test the studied geometry to contrast the theoretical results and, more importantly, the performance of the microfluidic-based recycling of the active medium.

\section{Acknowledgements}

This work was supported by Consejo Nacional de Investigaciones Cient\'ificas y T\'ecnicas (CONICET), Argentina; Secretar\'ia de Ciencia y Tecnolog\'ia de la Universidad Nacional de C\'ordoba (SECYT-UNC), Córdoba, Argentina; and Universidad de Buenos Aires, Project UBA 20020100100327, Buenos Aires, Argentina.

\bibliography{citeARXIV}

%\newpage
%\begin{figure*}[t]
%\includegraphics[width=7.0 in]{figure-TOC.pdf}
%\caption{\textbf{TOC graphic.}}
%\label{fig:TCO}
%\end{figure*}

\end{document}